\documentclass[11pt,twoside]{article}

\usepackage{asp2006}
\usepackage{epsfig}
\usepackage{graphicx}

\begin{document}
\title{Tidal tails around globular clusters}   
\author{M. Montuori$^{(1)}$,R. Capuzzo Dolcetta$^{(2)}$,P. Di Matteo$^{(3)}$,P. Miocchi$^{(2)}$}   
\affil{$^{(1)}$ CNR-INFM, Roma, Italia; $^{(2)}$ Dep. of Phys., Universit\'a di Roma «La Sapienza» , Roma, Italia; $^{(3)}$ Observatoire de Paris, LERMA, Paris, France}    

\begin{abstract} 

We present the results of detailed N-body simulations of 
clusters moving in a realistic Milky Way (MW) potential. 
The strong interaction with the bulge and the disk of the 
Galaxy leads to the formation of tidal tails, emanating 
from opposite sides of the cluster. Their orientation and morphology may be interpreted easily in terms of a comoving frame of coordinates.

\end{abstract}



\begin{figure} [h] 
\centering
\includegraphics[angle=0,width=11cm]{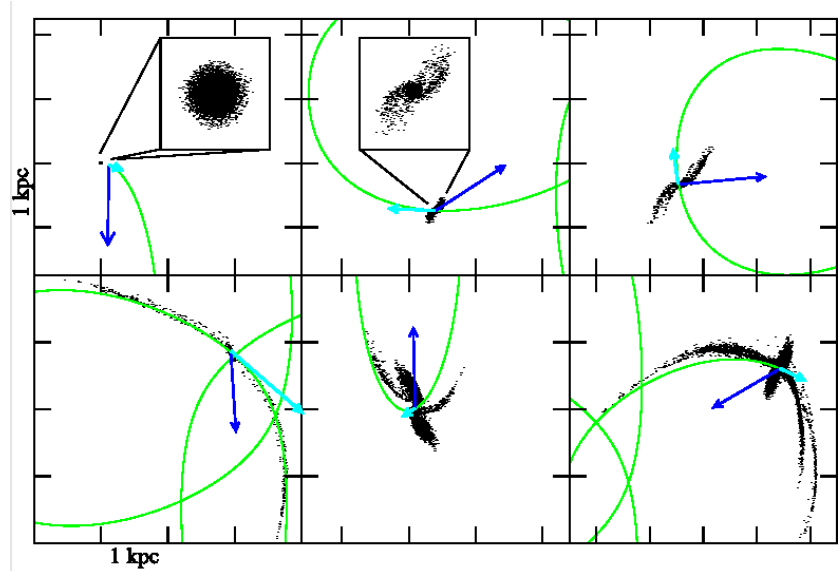}
\caption{Solid arrow is the GC velocity vector; dashed arrow points to the galactic center; 
the curve is the GC orbit.} 
\label{fig1} 
\end{figure}

Tidal tails are often observed around dwarf galaxies and globular clusters (GCs) \citep[see, e.g.][]{oden02}. Fig. \ref{fig1} shows the formation and evolution of tidal tails around a GC, simulated as an N-body system moving in MW potential \citep{mont07}. Tidal tails clearly show up after less than one orbital crossing. Their extension and orientation depend on the velocity and acceleration of the cluster along the orbit. The outer part (at distance $>$ 7-8 tidal radii) of the tails is aligned with the cluster orbit, while the inner part shows a shape which varies with time. When the cluster approaches its apogalacticon, tails divide into multiple ``arms'', as observed in NGC 288 and Willman 1.

\begin{figure} [h] 
\centering
\includegraphics[angle=0,width=12cm]{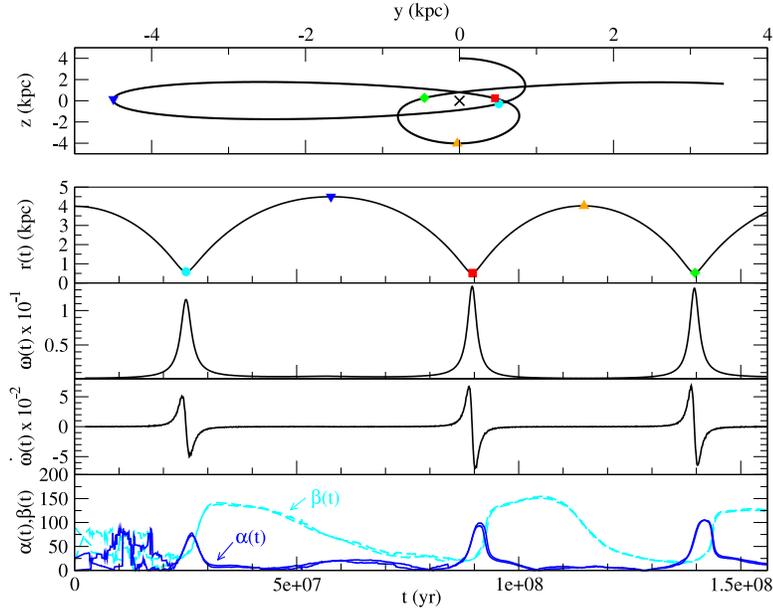}
\caption{Top down: 1st panel: GC orbit; 
2nd panel: GC distance to the galactic center; 3rd panel: GC angular velocity; 
4th panel: GC angular acceleration; 5th panel: angles $\alpha(t)$ and 
$\beta(t)$ between inner tail and radius vector r and angular velocity respectively. 
All the quantities shown in panels 2 to 5 are vs. time.}
\label{fig2} 
\end{figure}

Fig.\ref{fig2} shows the behaviour of some orbital quantities. Due to that the orientation of the inner tails is highly correlated to the cluster orbital phase and to the local orbital angular acceleration, the orbital path cannot be deduced directly from the orientation of the tails, unless a sufficient large field around the cluster is observed.


\end{document}